\documentstyle[12pt]{letter}
\begin{document}

\centerline{\bf{Cuprates: experiment vs. quasiparticle theory}}

Indeed, one can hardly disagree with most of the points raised by
P.W. Anderson in his Reference Frame of the February 2000 issue of
Physics Today. 

However, one can take a different viewpoint than his on the cuprates.

Experimentally, the 'notorious "linear T" resistivity' etc. has only
been observed  at temperatures T of the order of a few degrees Kelvin
{\em at most}, and {\em never} all the way down to (practically)
zero temperature. Somehow, many people in the field make the implicit
extrapolation of the linear regime down to T=0.
I think this is an important issue, which has been all too often 
sidestepped, especially in efforts to model the (normal state of the)
cuprates.

Now, regarding the irrelevance (or not!) of quasiparticle theory
thereon. Conceivably, ' "proofs" that purport to show otherwise,
are not very useful since they ignore the possibility of anomalies ...'.
{\em But,} are there experimental indications pointing to this possibility? 
It is easy to imagine that a long discussion may ensue on the matter.

Nevertheless, if one is willing to give good old quasiparticle perturbative
methods a chance,  the 'notorious "linear T" resistivity' and optical
conductivity can be {\em analytically} understood - at least for
overdoped and optimally doped materials. (Stripes apparently dominate the
physics of the underdoped regime, and a smooth crossover between
these regimes can be envisaged, but that is another story.) One obtains 
a one-particle scattering rate which goes like x$^2$, x=max\{T,energy\}, for 
x$\rightarrow$0. This scattering rate becomes {\em linear} in energy if the 
latter
exceeds a crossover value x$_o$, or {\em linear} in T if T exceeds (x$_o$/4).
(x$_o$ is the difference between the chemical potential and the energy
of the van Hove singularity.)
The result holds true {\em everywhere} in the Brillouin zone [1]. This 
prediction was directly supported by the ARPES expts. of Johnson et al. [2].
Experiment also indicates that x$_o$ is roughly in the range 50-350 $^o$K,
not in disagreement with the observed resistivity etc.

A couple more points related to traditional Fermi liquid effects in the
cuprates. In the model presented above, phonons play {\em no} essential role, 
if any, in the T dependence of the scattering rate. They merely provide momentum
dissipation, thus yielding a finite resistivity [1]. On the other hand, the
unconventional effects of impurity scattering, e.g. in the low T resistivity
[3], merit further investigation before any conclusions can be drawn.

\vspace{1cm}
George Kastrinakis

Dept. of Chemical Engineering\\
University of Cambridge \\
Cambridge CB2 3RA, U.K.

\vspace{1cm}

References

[1] G. Kastrinakis, Physica C {\bf 317-319}, 497 (1999); e-print 
cond-mat/0005485, Physica C (in print).

[2] T. Valla et al., Science {\bf 285}, 2110 (1999); e-print
cond-mat/0003407.

[3] G.S. Boebinger et al., Phys. Rev. Lett. {\bf 77}, 5417 (1996).
\end{document}